\documentclass[preprint,12pt]{elsarticle}

\usepackage{graphicx}
\usepackage{amssymb}
\usepackage{amsthm}

\usepackage{amsmath}
\usepackage{subfigure}

\journal{Computer Physics Communications}

\begin{document}

\begin{frontmatter}



\title{Event-based simulation of light propagation in lossless
dielectric media\footnote{Accepted for publication in Computer Physics Reports}}


\author[FZJ]{Binh Trieu}
\author[FZJ]{Kristel Michielsen}
\author[RUG]{Hans De Raedt}

\address[FZJ]{
Institute for Advanced Simulation, J\"ulich Supercomputing Centre,
Research Centre J\"ulich, D-52425 J\"ulich, Germany}
\address[RUG]{Department of Applied Physics, Zernike Institute for Advanced Materials,
University of Groningen, Nijenborgh 4, NL-9747 AG Groningen, The Netherlands}

\begin{abstract}
We describe an event-based approach to simulate the
propagation of an electromagnetic plane wave through dielectric
media. The basic building block is a deterministic learning machine
that is able to simulate a plane interface. We show that
a network of two of such machines can simulate the propagation of light through a
plane parallel plate. With properly chosen parameters this setup can
be used as a beam splitter. The modularity of the
simulation method is illustrated by constructing a Mach-Zehnder interferometer
from plane parallel plates, the whole system reproducing the results
of wave theory. A generalization of the event-based model of the plane parallel plate is also used to
simulate a periodically stratified medium.
\end{abstract}


\end{frontmatter}


\section{Introduction}
\label{intro}

Maxwell's theory of electrodynamics forms the basis
of the understanding of the properties of light~\cite{BORN64}.
The Maxwell equations describe the evolution of electromagnetic fields in space and time~\cite{BORN64}.
They apply to a wide range of different physical situations and play an important role in a large number of engineering
applications. Maxwell's theory describes physical phenomena in terms of waves of electromagnetic radiation,
yielding a simple explanation for the observation of interference phenomena.
For many applications, computer simulation methods are required to solve Maxwell's equations, the
work horse being the finite-difference time-domain (FDTD) method~\cite{TAFL05}.

In this paper, we present an alternative to the FDTD method.
In contrast to a wave-based description, our approach uses
particles (photons) that interact with matter.
The simulation proceeds event-by-event, that is particle-by-particle.
There is no direct communication/interaction between different particles:
Indirect communication takes place via the interaction with matter,
an interaction that is modeled by means of
a deterministic learning machine (DLM)~\cite{RAED05b,RAED05c,RAED05d}.
As we show in the paper, our approach is modular and
yields the same stationary-state results as those obtained from Maxwell's theory.

In section~\ref{interface}, we introduce the simulation approach and
explain how it can be used to describe the reflection properties of a single
interface, including interference effects
(section~\ref{interface_interference}).
The modularity of our approach is illustrated by combining two or more DLMs
to describe a homogeneous dielectric film (section~\ref{plate}) and a
multilayer (section~\ref{multilayer}).
Finally, we show that the basic building blocks can be re-used
without modification to simulate more complex optical devices
such as a Mach-Zehnder interferometer (MZI) (section~\ref{mach-zehnder}).

\section{Reflection and refraction at an interface}
\label{interface}
\subsection{Wave theory}
\begin{figure}
\begin{center}
\includegraphics[width=10cm]{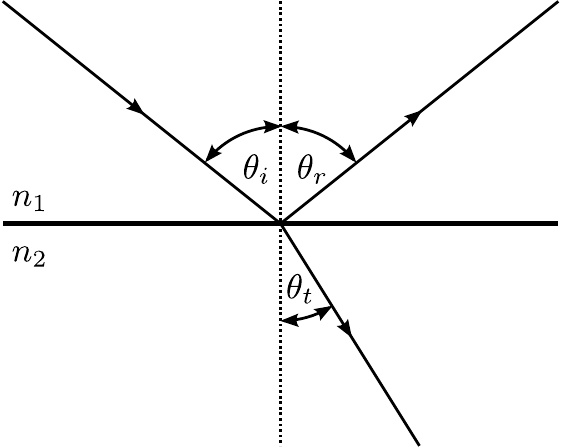}
\end{center}
\caption{Reflection and refraction of light at a plane interface in
  the plane of incidence. The angle of refraction $\theta_t$ is
  determined by the angle of incidence $\theta_i$ and the refractive
  indices of both media $n_1$ and $n_2$ (Eq.~\ref{eq:Snells}). The
  angle of reflection $\theta_r$ is equal to the angle of incidence
  $\theta_i$.}
\label{fig:interface}
\end{figure}
In classical electrodynamics the laws of reflection and refraction are
derived from Maxwell's theory. In case of a plane wave incident on an
interface between two homogeneous isotropic media with different
optical properties, there is in general a transmitted wave and a
reflected wave. The angle of incidence and the refractive indices of
both media determine the direction of the transmitted and reflected
part. Figure~\ref{fig:interface} shows a schematic picture in the
plane of incidence. The relation between the angle of the incident
ray $\theta_i$ and that of the reflected ray $\theta_r$ is determined
by the law of reflection,
\begin{equation}
\theta_r=\theta_i.
\end{equation}
The direction of the transmitted wave is determined by the law of
refraction,
\begin{equation}
\frac{\sin\theta_i}{\sin\theta_t}=\frac{n_2}{n_1},
\label{eq:Snells}
\end{equation}
where $\theta_t$ is the angle of the transmitted ray, $n_1$ is the
refractive index of the first and $n_2$ is that of the second
medium (see Fig.~\ref{fig:interface}).

For lossless, perfectly transparent media, the wave
amplitudes are given by the Fresnel formulae~\cite{BORN64}:
\begin{eqnarray}
\label{eq:fresnel_formulae}
T_\parallel&=&\frac{2n_1\cos\theta_i}{n_2\cos\theta_i+n_1\cos\theta_t}
A_\parallel,\\
\label{eq:fresnel_formulae2}
T_\perp&=&\frac{2n_1\cos\theta_i}{n_1\cos\theta_i+n_2\cos\theta_t}A_\perp,\\
\label{eq:fresnel_formulae3}
R_\parallel&=&\frac{n_2\cos\theta_i-n_1\cos\theta_t}
{n_2\cos\theta_i+n_1\cos\theta_t}A_\parallel,\\
\label{eq:fresnel_formulae4}
R_\perp&=&\frac{n_1\cos\theta_i-n_1\cos\theta_t}{n_1\cos\theta_i+n_2\cos\theta_t}A_\perp.
\end{eqnarray}
The components of the electric field vector of the incident
electromagnetic field parallel and perpendicular to the plane of incidence
are denoted by $A_\parallel$ and $A_\perp$, respectively.
$T_\parallel$ and $T_\perp$ are the amplitudes of the transmitted wave and $R_\parallel$
and $R_\perp$ are the amplitudes of the reflected wave.

The reflectivity $\mathcal{R}$ and the transmissivity $\mathcal{T}$
are given by
\begin{equation}
\label{eq:reflectivity}
\mathcal{R}=\frac{|R|^2}{|A|^2}
,
\end{equation}
and
\begin{equation}
\label{eq:transmissivity}
\mathcal{T}=\frac{n_2}{n_1}\frac{\cos{\theta_t}}{\cos{\theta_i}}
\frac{|T|^2}{|A|^2}.
\end{equation}
Equations~(\ref{eq:reflectivity}) and (\ref{eq:transmissivity}) are
both valid for the parallel as well as for the perpendicular part and both
components can be treated separately. Combining
Eqs.~(\ref{eq:fresnel_formulae}) to (\ref{eq:transmissivity}) and
simplifying the expressions gives~\cite{BORN64}
\begin{eqnarray}
\mathcal{R}_\parallel&=&\frac{\tan^2(\theta_i-\theta_t)}
{\tan^2(\theta_i+\theta_t)},
\label{eq:RT1}
\\
\mathcal{R}_\perp&=&\frac{\sin^2(\theta_i-\theta_t)}
{\sin^2(\theta_i+\theta_t)},\\
\mathcal{T}_\parallel&=&\frac{\sin2\theta_i\sin2\theta_t}
{\sin^2(\theta_i+\theta_t)\cos^2(\theta_i-\theta_t)},\\
\mathcal{T}_\perp&=&\frac{\sin2\theta_i\sin2\theta_t}{\sin^2(\theta_i+\theta_t)}.
\label{eq:RT4}
\end{eqnarray}

\subsection{Event-based simulation}
We simulate the behavior of reflection and transmission at an
interface by using an event-by-event, particle-only approach~\cite{RAED05b,RAED05c,RAED05d}. 
Clearly, in an event-based model of refraction and reflection at a
dielectric, lossless interface, there can be no loss of particles:
An incident particle must either bounce back from
or pass through the interface. If such a model is to reproduce
the results of Maxwell's theory, the boundary
conditions on the wave amplitudes in Maxwell's theory
must translate into a rule that determines how a particle
bounces back or crosses the interface.
In this section, we specify these rules.

We call an event the arrival of a single photon at the interface. This
photon carries a message that can be interpreted as phase or
time-of-flight information. As events occur one at a time only, there
is no communication between individual photons, but the exchange and
the processing of information takes place within the apparatus that
describes the interface. An incoming photon will either be reflected
or transmitted, depending on the state of the processing unit.

\begin{figure}
\begin{center}
\includegraphics[width=10cm]{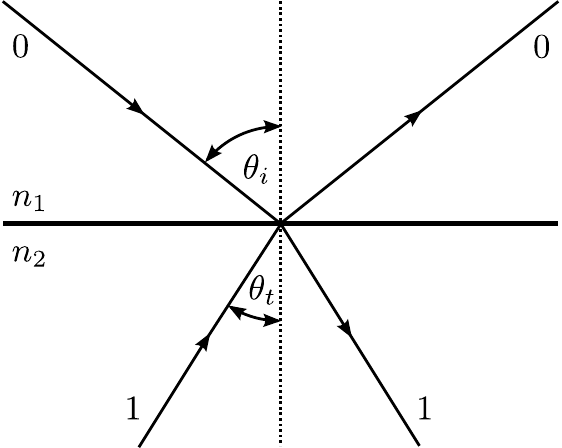}
\end{center}
\caption{Reflection and refraction of light at a plane interface in
  the plane of incidence. There are two input ports (0 and 1)
  and two output ports (0 and 1). The angles are chosen such that the
  direction of a refracted outgoing photon coincides with the
  direction of a reflected photon of the opposite input port
  and vice versa, allowing for interference to occur.}
\label{fig:interface2}
\end{figure}

For later use, in addition to the input port and two output ports
as depicted in Fig.~\ref{fig:interface}, we add an additional
input port that captures light incident from the opposite direction in such a way
that the direction of a refracted outgoing particle coincides with the
direction of a reflected particle of the opposite input port and
vice versa (see Fig.~\ref{fig:interface2}).

The processing unit in this case is a DLM~\cite{RAED05b,RAED05c,RAED05d} 
that takes messages from two input ports and sends out messages on
either of two possible output ports depending on the internal
state. The internal state is updated with each message that the DLM
receives, i.e.\ it learns from the events that it processes.

\begin{figure}
\begin{center}
\includegraphics[width=\textwidth]{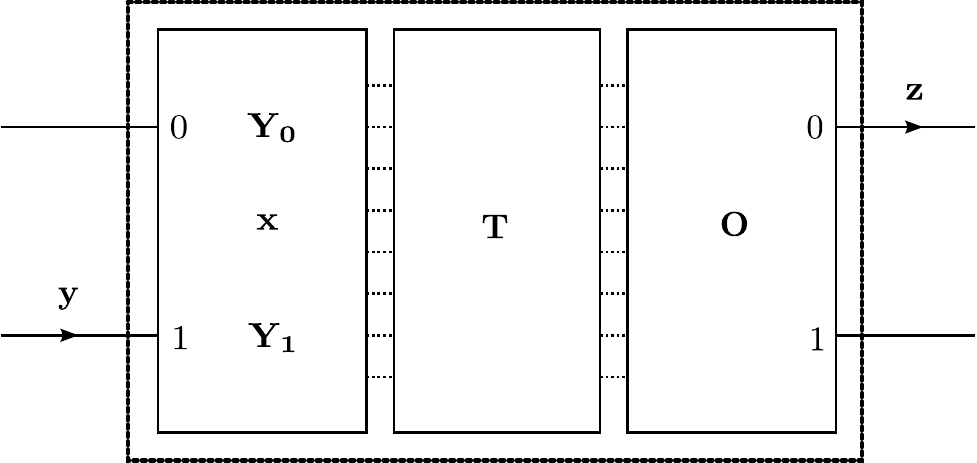}
\end{center}
\caption{Schematic diagram of a DLM that performs an event-based
  simulation of a plane interface. There are two input ports and
  two output ports. The first stage (DLM) updates the internal registers
  $\mathbf{Y}_0$, $\mathbf{Y}_1$ and $\mathbf{x}$ according to the input
  $\mathbf{y}$. The second stage (T) processes the information stored in
  these registers according to a specific rule (Eq.~(\ref{eq:T})) and the third stage (O)
  prepares the outgoing message and sends it through one of the output ports.}
\label{fig:dlm}
\end{figure}

The deterministic learning machine consists of three stages (see
Fig.~\ref{fig:dlm}): The first stage receives an input from the
n\textsuperscript{th} event, in this case the phase information
$\phi$ of the photon and the angle of polarization $\varpi$, for practical
reasons encoded as a four-dimensional vector
$\mathbf{y}_n=((y_n)_{0,\parallel},(y_n)_{1,\parallel},
(y_n)_{0,\perp},(y_n)_{1,\perp}),$
with
$(y_n)_{0,\parallel}=\cos(\phi)\cos(\varpi)$,
$(y_n)_{1,\parallel}=\sin(\phi)\cos(\varpi)$,
$(y_n)_{0,\perp}=\cos(\phi)\sin(\varpi)$, and
$(y_n)_{1,\perp}=\sin(\phi)\sin(\varpi)$.
Upon arrival of a photon at one of its two input ports,
the DLM stores the message in its internal register $\mathbf{Y}_k$ with $k=0$
or $k=1$ if the input was on port 0 or 1, respectively. There is
also an internal vector
$\mathbf{x}=(x_0, x_1)$ with $x_i \in [0,1], i=\{0,1\}$ and
$x_0+x_1=1$. This vector is updated for each event received on port
$k$ according to
\begin{equation}
(x_{n+1})_i=\alpha (x_n)_i + (1-\alpha)\delta_{i,k},
\end{equation}
where $0 < \alpha < 1$ is a parameter that determines the speed of
learning. Since $(x_n)_0+(x_n)_1=1$ for all $n$, we can interpret
$(x_n)_k$ as (an estimate of) the frequency for the occurrence of an
event on port $k$.

The second stage processes the information stored in the registers
$\mathbf{Y}_0, \mathbf{Y}_1$ and $\mathbf{x}$ according to the rule
\begin{equation}
\mathbf{T}
\left(\begin{array}{c}
(Y_0)_{0,\parallel}\sqrt{x_0}\\
(Y_0)_{1,\parallel}\sqrt{x_0}\\
(Y_1)_{0,\parallel}\sqrt{x_1}\\
(Y_1)_{1,\parallel}\sqrt{x_1}\\
(Y_0)_{0,\perp}\sqrt{x_0}\\
(Y_0)_{1,\perp}\sqrt{x_0}\\
(Y_1)_{0,\perp}\sqrt{x_1}\\
(Y_1)_{1,\perp}\sqrt{x_1}\\
\end{array}\right)
=
\left(\begin{array}{c}
r_\parallel(Y_0)_{0,\parallel}\sqrt{x_0}+t_\parallel(Y_1)_{0,\parallel}\sqrt{x_1}\\
r_\parallel(Y_0)_{1,\parallel}\sqrt{x_0}+t_\parallel(Y_1)_{1,\parallel}\sqrt{x_1}\\
t_\parallel(Y_0)_{0,\parallel}\sqrt{x_0}-r_\parallel(Y_1)_{0,\parallel}\sqrt{x_1}\\
t_\parallel(Y_0)_{1,\parallel}\sqrt{x_0}-r_\parallel(Y_1)_{1,\parallel}\sqrt{x_1}\\
r_\perp(Y_0)_{0,\perp}\sqrt{x_0}+t_\perp(Y_1)_{0,\perp}\sqrt{x_1}\\
r_\perp(Y_0)_{1,\perp}\sqrt{x_0}+t_\perp(Y_1)_{1,\perp}\sqrt{x_1}\\
t_\perp(Y_0)_{0,\perp}\sqrt{x_0}-r_\perp(Y_1)_{0,\perp}\sqrt{x_1}\\
t_\perp(Y_0)_{1,\perp}\sqrt{x_0}-r_\perp(Y_1)_{1,\perp}\sqrt{x_1}\\
\end{array}\right),
\label{eq:T}
\end{equation}
with
\begin{eqnarray}
\label{eq:r_parallel}
r_\parallel&=&\sqrt{\mathcal{R}_\parallel},\\
\label{eq:t_parallel}
t_\parallel&=&\sqrt{\mathcal{T}_\parallel},\\
\label{eq:r_perp}
r_\perp&=&\sqrt{\mathcal{R}_\perp},\\
\label{eq:t_perp}
t_\perp&=&\sqrt{\mathcal{T}_\perp}.
\end{eqnarray}
Here we have omitted the event label $n$.

The third stage of the DLM prepares the messages
\begin{equation}
\mathbf{w}_n=\left(\begin{array}{c}
t_\parallel(Y_0)_{0,\parallel}\sqrt{x_0}-r_\parallel(Y_1)_{0,\parallel}\sqrt{x_1}\\
t_\parallel(Y_0)_{1,\parallel}\sqrt{x_0}-r_\parallel(Y_1)_{1,\parallel}\sqrt{x_1}\\
t_\perp(Y_0)_{0,\perp}\sqrt{x_0}-r_\perp(Y_1)_{0,\perp}\sqrt{x_1}\\
t_\perp(Y_0)_{1,\perp}\sqrt{x_0}-r_\perp(Y_1)_{1,\perp}\sqrt{x_1}\\
\end{array}\right)
,
\end{equation}
and
\begin{equation}
\mathbf{w}'_n=\left(\begin{array}{c}
r_\parallel(Y_0)_{0,\parallel}\sqrt{x_0}+t_\parallel(Y_1)_{0,\parallel}\sqrt{x_1}\\
r_\parallel(Y_0)_{1,\parallel}\sqrt{x_0}+t_\parallel(Y_1)_{1,\parallel}\sqrt{x_1}\\
r_\perp(Y_0)_{0,\perp}\sqrt{x_0}+t_\perp(Y_1)_{0,\perp}\sqrt{x_1}\\
r_\perp(Y_0)_{1,\perp}\sqrt{x_0}+t_\perp(Y_1)_{1,\perp}\sqrt{x_1}\\
\end{array}\right)
,
\end{equation}
and generates a uniform random number $0 < r < 1$. If $\|\mathbf{w}_n\| >
r$, the final stage sends $\mathbf{z}_n=\mathbf{w}_n/\|\mathbf{w}_n\|$ through port 0.
Otherwise it sends $\mathbf{z}'_n=\mathbf{w}'_n/\|\mathbf{w}'_n\|$ through port 1.

From the above construction of the DLM, it is clear that the connection between
Maxwell's wave description and the event-based, particle-like
simulation model enters through Eqs.~(\ref{eq:r_parallel})-(\ref{eq:t_perp}),
where the expressions in the left-hand-sides of the latter are given by
Eqs.~(\ref{eq:RT1})-(\ref{eq:RT4}).

\subsection{Simulation results}
\begin{figure}
\begin{center}
\includegraphics[width=\textwidth]{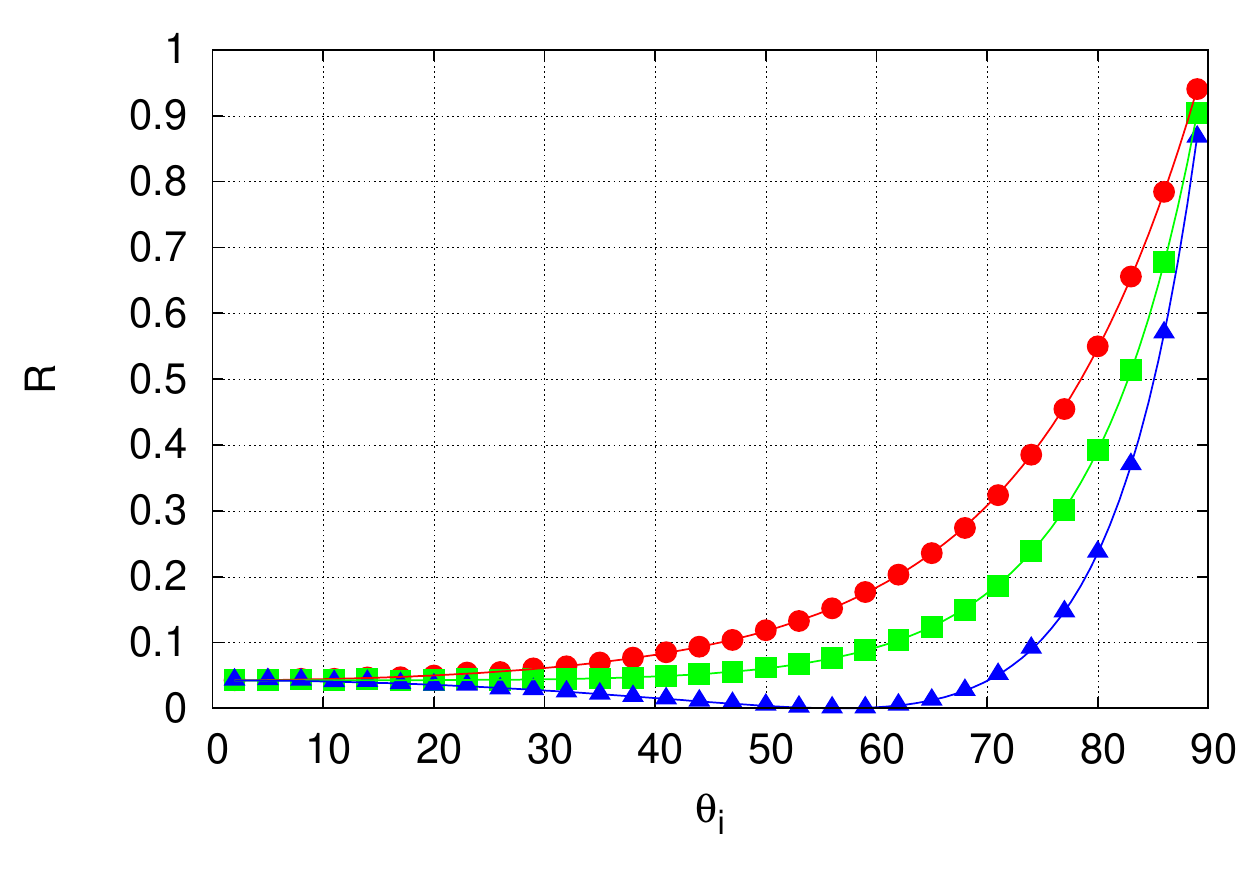}
\end{center}
\caption{Reflectivity $\mathcal{R}$ of a plane interface between
  homogeneous dielectric media with refractive indices $n_1=1$ and
  $n_2=1.52$ as a function of the angle of incidence $\theta_i$.
  Triangles: Polarization angle $\varpi=0$;
  Squares: $\varpi=\pi/4$;
  Circles: $\varpi=\pi/2$.
  Each marker represents a simulation of 100000 events with $\alpha=0.99$.
  The simulation data are in very good agreement with the wave
  theoretical predictions (solid lines).
  }
\label{fig:single_interface}
\end{figure}

As a first validation of the simulation model,
we simulate a single interface with our event-based method. The
initial values of the registers $\mathbf{Y}_0, \mathbf{Y}_1$ and $\mathbf{x}$
are chosen randomly, but properly normalized. For each point in
Fig.~\ref{fig:single_interface}, 100000 events were simulated by
sending messages on port 0 with a randomly chosen but fixed
phase. The parameters were set to $\alpha=0.99$, $n_1=1$, and
$n_2=1.52$. At the end we count, how many events we have detected on
output port 0, i.e.\ the fraction of reflected particles. This
normalized intensity corresponds to the reflectivity $\mathcal{R}$.

\subsection{Discussion}
We have shown that the reflectivity of a single interface can be
simulated by our event-based approach. The simulation results are in excellent
agreement with the theoretical predictions of Maxwell's wave theory.
Features such as the polarization dependence or reflection at the Brewster
angle are faithfully reproduced.
However, this good agreement does not show yet that our
model correctly simulates interference phenomena.
Such a demonstration is given in section~\ref{interface_interference}.

\section{Interference effects at an interface}
\label{interface_interference}

In section~\ref{interface}, we dealt with the case of
input on a single input port only.
Here we consider the case where particles can arrive,
one-by-one, on both sides of the interface, that is on both input ports.

\subsection{Wave theory}

According to Maxwell's theory, if particles enter on input port 0 with a probability of $p_0$,
carrying a phase $\phi_0$ and on input port 1 with probability
$1-p_0$ and phase $\phi_1$ the amplitudes on the output ports are given by
\begin{equation}
\left(\begin{array}{c}
b_0\\
b_1
\end{array}\right)
=\left(\begin{array}{cc}
r_{\{\parallel,\perp\}} & t_{\{\parallel,\perp\}}\\
t_{\{\parallel,\perp\}} & -r_{\{\parallel,\perp\}}
\end{array}\right)
\left(\begin{array}{c}
\sqrt{p_0}e^{i\phi_0}\\
\sqrt{1-p_0}e^{i\phi_1}
\end{array}\right),
\end{equation}
with $r_{\{\parallel,\perp\}}$ and $t_{\{\parallel,\perp\}}$ given by
Eqs.~(\ref{eq:r_parallel}) to (\ref{eq:t_perp}).
For S-polarization, i.e.\ the $\perp$-component ($\varpi=\pi/2$), the
normalized intensity on output port 0 or the reflectivity
$\mathcal{R}_\perp$ is given by
\begin{multline}
|b_{0,\perp}|^2 =
\frac{\sin^2(\theta_i-\theta_t)}{\sin^2(\theta_i+\theta_t)}p_0
+\frac{\sin2\theta_i\sin2\theta_t}{\sin^2(\theta_i+\theta_t)}(1-p_0)\\
-2\frac{\sin(\theta_i-\theta_t)\sqrt{\sin 2\theta_i\sin2\theta_t}\cos(\phi_0-\phi_1)}{\sin^2(\theta_i+\theta_t)}\sqrt{p_0(1-p_0)}.
\label{eq:interface_2channels_s}
\end{multline}
For P-polarization ($\varpi=0$), the corresponding expression reads
\begin{multline}
\label{eq:interface_2channels_p}
|b_{0,\parallel}|^2=\frac{\tan^2(\theta_i-\theta_t)}{\tan^2(\theta_i+\theta_t)}p_0
+\frac{\sin2\theta_i\sin2\theta_t}{\sin^2(\theta_i+\theta_t)\cos^2(\theta_i-\theta_t)}(1-p_0)\\
+2\frac{\tan(\theta_i-\theta_t)\sqrt{\sin2\theta_i\sin2\theta_t}\cos(\phi_0-\phi_1)
}{\tan(\theta_i+\theta_t)\sin(\theta_i+\theta_t)\cos(\theta_i-\theta_t)}\sqrt{p_0(1-p_0)}.
\end{multline}

\subsection{Simulation results}
\begin{figure}
\subfigure[]{\includegraphics[width=0.49\textwidth]{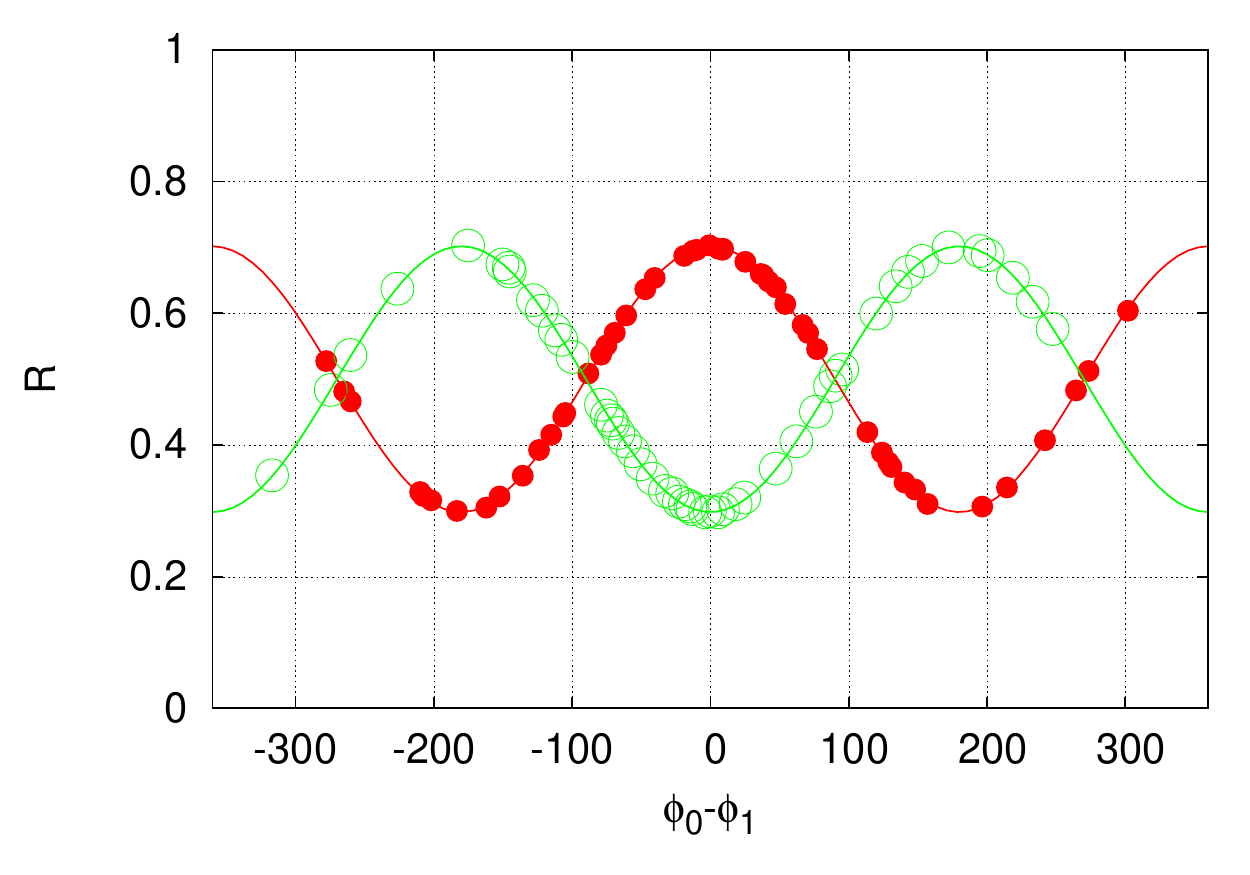}}
\subfigure[]{\includegraphics[width=0.49\textwidth]{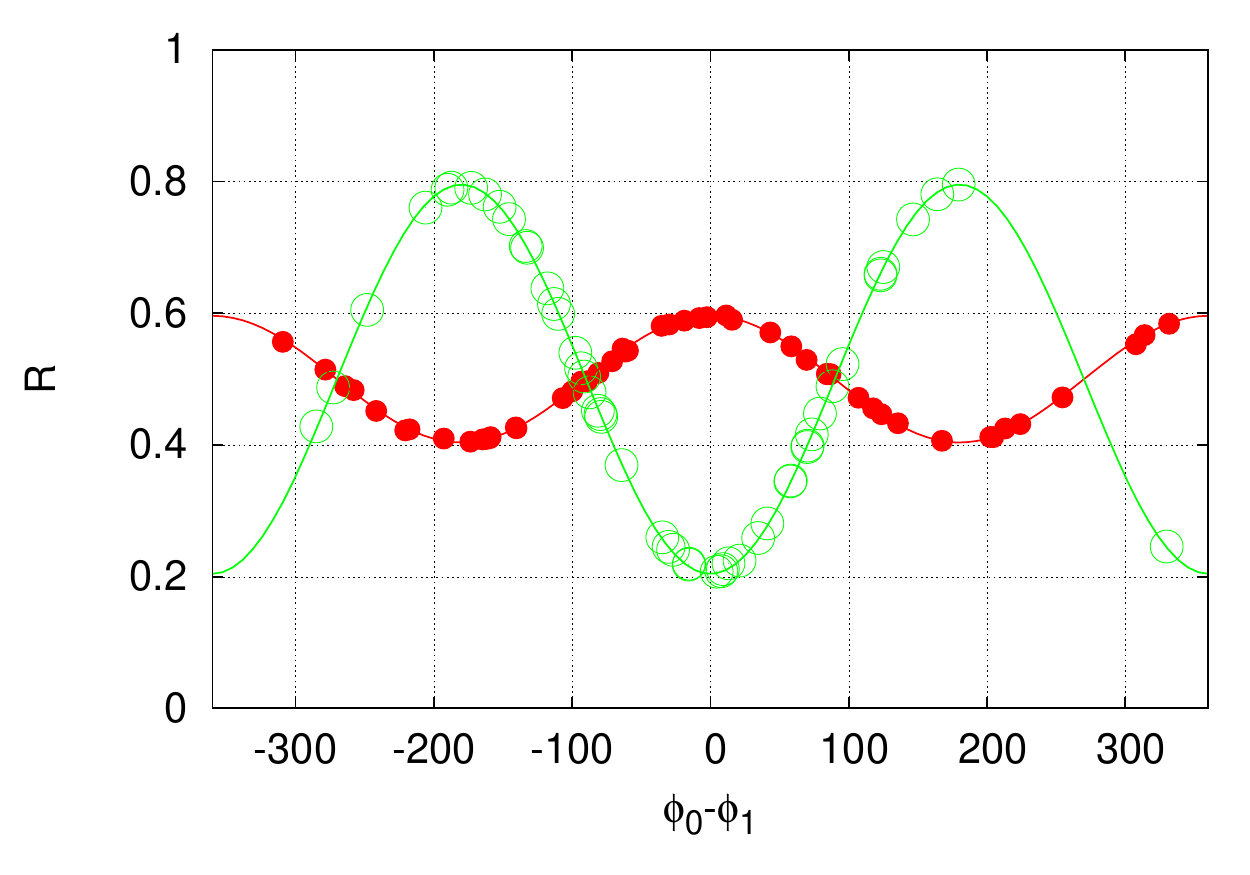}}
\subfigure[]{\includegraphics[width=0.49\textwidth]{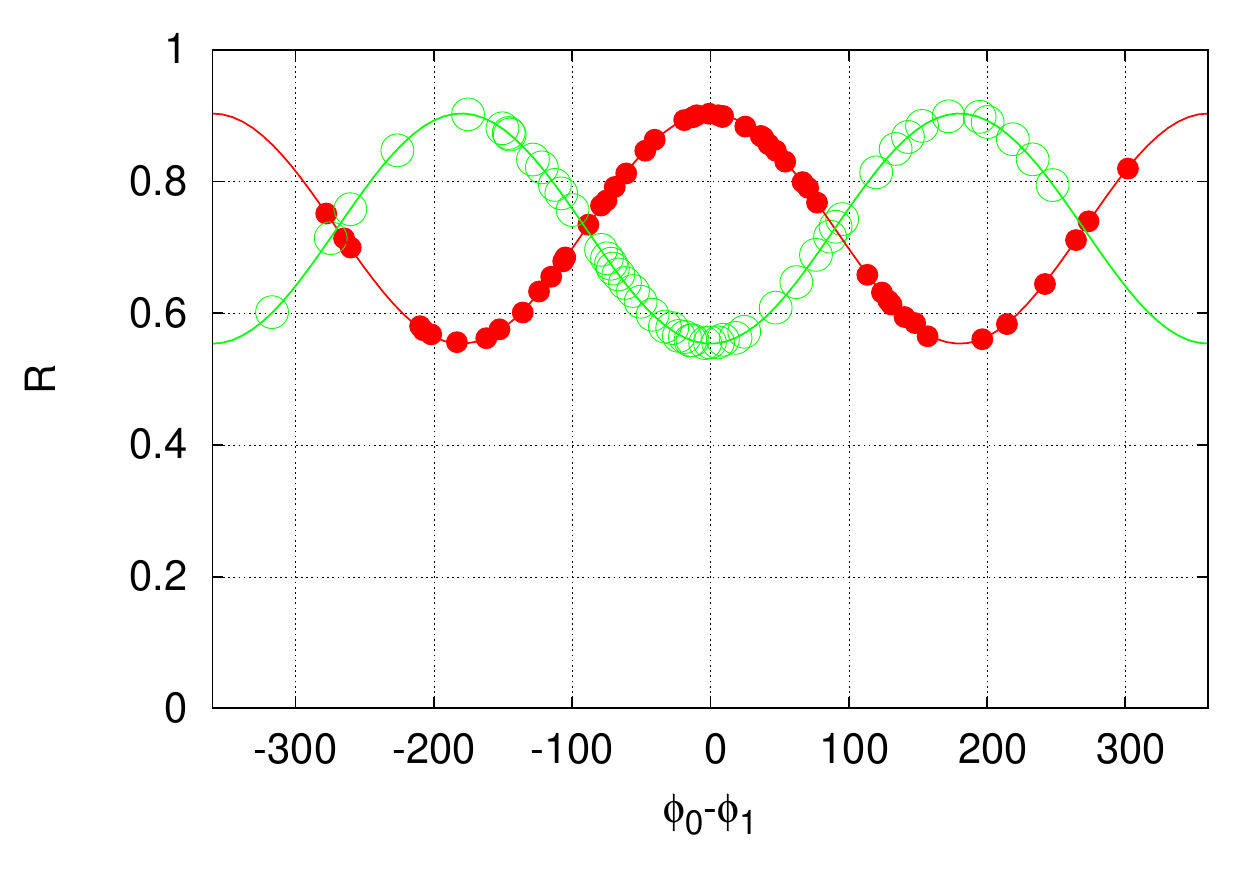}}
\subfigure[]{\includegraphics[width=0.49\textwidth]{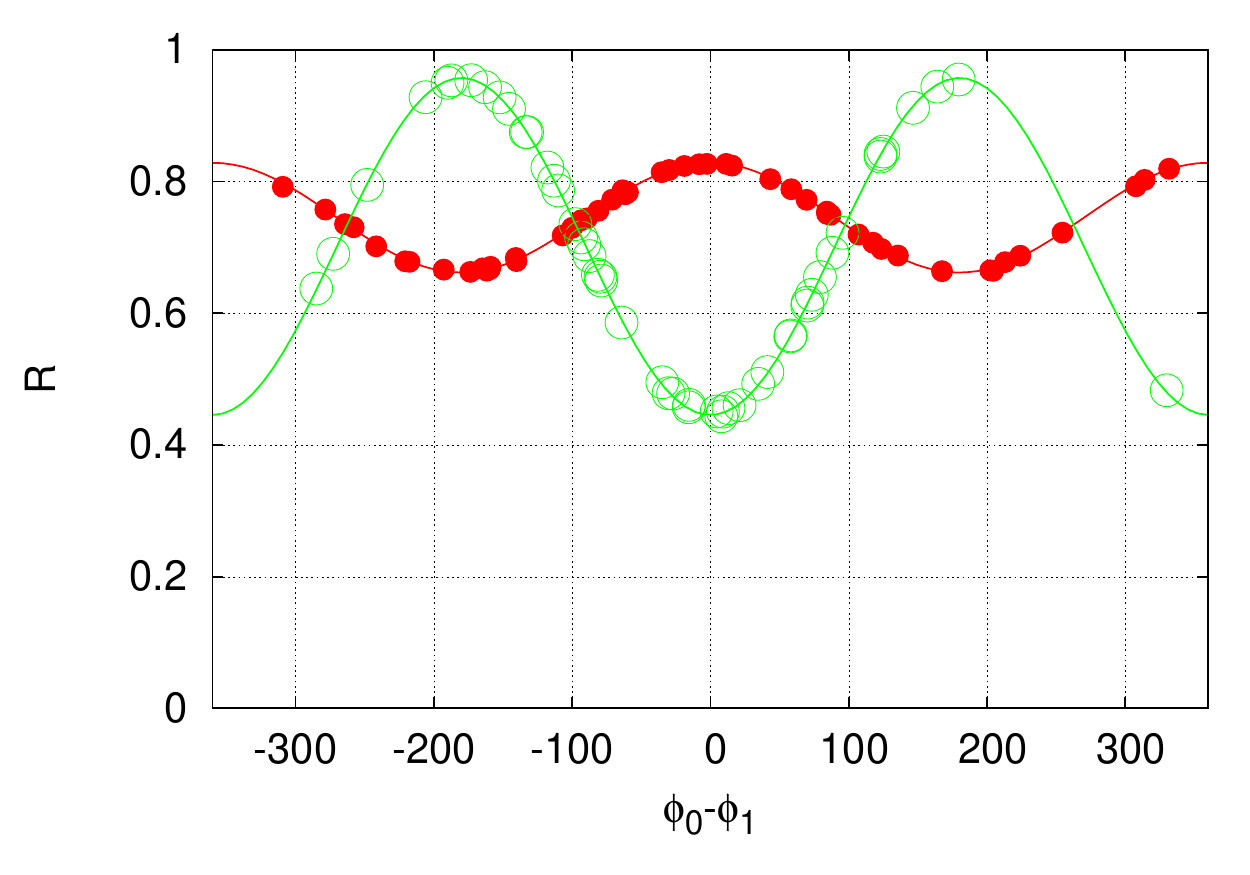}}
\caption{Single interface between a $n_1=1$ and a $n_2=1.52$ dielectric with input from both sides.
Single photons arrive on port 0 with probability $p_0$, carrying a phase
  $\phi_0$ and on port 1 with probability $1-p_0$ and phase
  $\phi_1$. At any time there is at most one single photon in the system.
  Each marker represents a simulation of 100000 events with $\alpha=0.98$.
  The phase difference $\phi_0-\phi_1$ is generated from randomly chosen
  values of $\phi_0$ and $\phi_1$ (which are then fixed for 100000 events).
  Solid circles: Polarization angle $\varpi=0$;
  Open circles: $\varpi=\pi/2$.
  (a): $p_0=1/2$, $\theta_i=0$;
  (b): $p_0=1/2$, $\theta_i=\pi/4$;
  (c): $p_0=1/4$, $\theta_i=0$;
  (c): $p_0=1/4$, $\theta_i=\pi/4$.
  The simulation results are in
  excellent agreement with the theoretical expressions derived from
  wave theory (Eqs.~(\ref{eq:interface_2channels_s}) and
  (\ref{eq:interface_2channels_p})), shown here as solid lines.}
 \label{fig:single_interface_input2}
\end{figure}

In Fig.~\ref{fig:single_interface_input2} we compare the event-based 
simulation results to the wave theoretical predictions Eqs.~(\ref{eq:interface_2channels_s}) and (\ref{eq:interface_2channels_p}).
In these simulations, we send, one-by-one,
photons with phase $\phi_0$ on port 0 with probability $p_0$ and
photons with phase $\phi_1$ on port 1 with probability $1-p_0$.
Of course, the results depend on the incident angle $\theta_i$.

\subsection{Discussion}
Our event-based simulation results are in
excellent agreement with the wave theoretical description.
For a single interface and a single input port there are no
interference effects and the results are in agreement with wave theory,
independent of the parameter $\alpha$.
However, in the case of two input ports where interference can occur,
the value of $\alpha$ is important:
The wave theoretical predictions can only be
reproduced by the event-based simulation if $\alpha$ is close to one~\cite{RAED05b,RAED05c,RAED05d}.

\section{Light propagation through a homogeneous dielectric film
  (plane-parallel plate)}
\label{plate}

\begin{figure}
\begin{center}
\includegraphics[width=8cm]{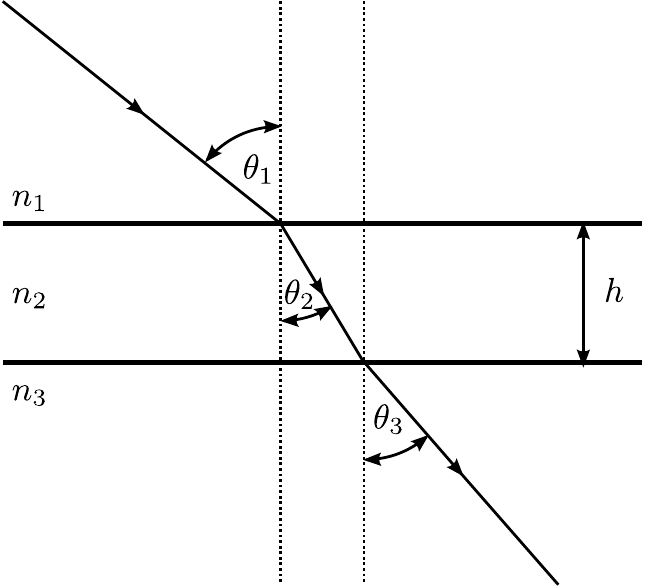}
\end{center}
\caption{Homogeneous dielectric film built from two plane parallel
  interfaces. The film thickness is denoted by $h$. For a plate surrounded by air the
  refractive indices $n_1=n_3=1$ and  $\theta_1=\theta_3$.}
\label{fig:plate}
\end{figure}

A homogeneous dielectric film between two homogeneous media can be
regarded as two plane parallel interfaces. 
Figure~\ref{fig:plate} shows a schematic picture of the system.

\subsection{Wave theory}
According to Maxwell's theory, the reflectivity and transmissivity of the film are given by~\cite{BORN64}
\begin{equation}
\mathcal{R}=\frac{r_{12}^2+r_{23}^2+2r_{12}r_{23}\cos2\beta}{1+r_{12}^2r_{23}^2+2r_{12}r_{23}\cos2\beta},
\label{eq:plate_reflectivity}
\end{equation}
and
\begin{equation}
\mathcal{T}=\frac{n_3\cos\theta_3}{n_1\cos\theta_1}\frac{t_{12}^2t_{23}^2}{1+r_{12}^2r_{23}^2+2r_{12}r_{23}\cos2\beta},
\end{equation}
with
\begin{equation}
r_{12}=\frac{n_1\cos\theta_1-n_2\cos\theta_2}{n_1\cos\theta_1+n_2\cos\theta_2},
\label{eq:plate_r}
\end{equation}
\begin{equation}
t_{12}=\frac{2n_1\cos\theta_1}{n_1\cos\theta_1+n_2\cos\theta_2},
\label{eq:plate_t}
\end{equation}
for a S-polarized wave ($\varpi=\pi/2$) and
\begin{equation}
r_{12}=\frac{n_2\cos\theta_1-n_1\cos\theta_2}{n_2\cos\theta_1+n_1\cos\theta_2},
\label{eq:plate_r_p}
\end{equation}
\begin{equation}
t_{12}=\frac{2n_1\cos\theta_1}{n_2\cos\theta_1+n_1\cos\theta_2},
\label{eq:plate_t_p}
\end{equation}
for a P-polarized wave ($\varpi=0$),
describing the process at the interface from medium 1 to medium 2 and
analogous expressions for $r_{23}$ and $t_{23}$. The variable $\beta$
is determined by
\begin{equation}
\beta=\frac{2\pi}{\lambda_0}n_2h\cos\theta_2,
\label{eq:beta}
\end{equation}
with $\lambda_0$ being the wavelength of the incident wave and $h$
denoting the thickness of the film.

\subsection{Event-based simulation}
We simulate a homogeneous dielectric film by connecting two DLMs, one for
each interface and each working as described in section~\ref{interface}.
For each interface, we use the appropriate expressions
for $r$ and $t$ as given by Eqs.~(\ref{eq:r_parallel})--(\ref{eq:t_perp})
and Eqs.~(\ref{eq:RT1})--(\ref{eq:RT4}), that is we do {\bf not}
use the expressions from Maxwell's theory for the film
(Eqs.~(\ref{eq:plate_reflectivity})--(\ref{eq:beta})).
We recover the results of Maxwell's theory by
the event-based simulation without solving the wave equation for the film.
This modularity allows us to re-use the DLM model
of an interface for simulating films, multilayers etc.

\begin{figure}
\begin{center}
\includegraphics[width=\textwidth]{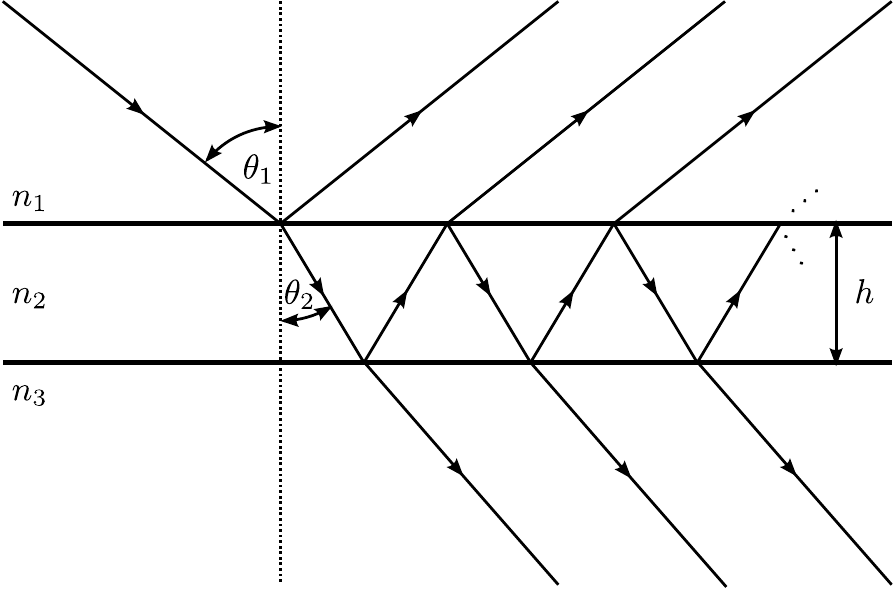}
\end{center}
\caption{A homogeneous dielectric film built from two plane parallel
  interfaces. Upon incidence on either of the two interfaces a photon
  can either be reflected or refracted. This can continue until it
  leaves on any side of the film. Due to the translational invariance
  when dealing with plane waves, corresponding translated paths can be
  regarded as superimposed.}
\label{fig:film}
\end{figure}

\begin{figure}
\begin{center}
\includegraphics[width=\textwidth]{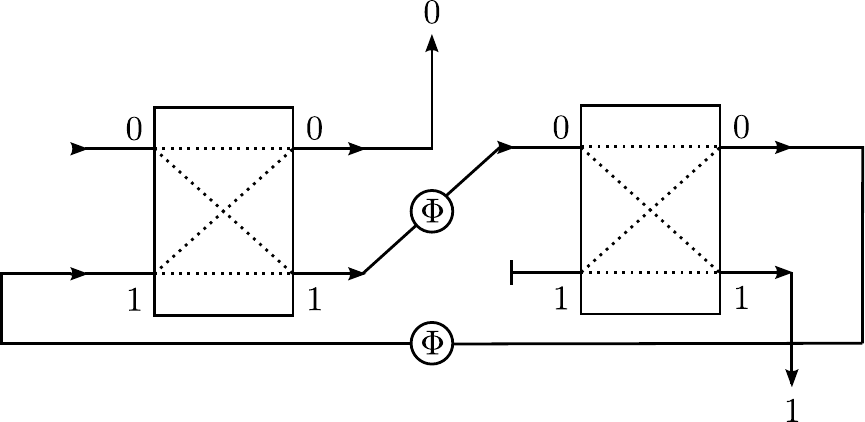}
\end{center}
\caption{Arrangement of two DLMs simulating a homogeneous dielectric
  film. The output on port 0 of the second DLM is fed back into
  the input port 1 of the first machine. The symbol $\Phi$ denotes
  a phase shift that a message acquires when travelling that
  specific path.}
\label{fig:film_dlm}
\end{figure}

We assume that the path of the particles is as depicted
in Fig.~\ref{fig:film}. The incident particle is either reflected or
refracted as it hits the first interface. In case of reflection it
leaves the system on the front side of the film. If it is
refracted, the particle refracted from the first interface
acts as the incident particle of the second interface. While traveling from the
first to the second interface it acquires a phase shift $\exp(i\Phi)$,
with $\Phi=\beta$ (Eq.~(\ref{eq:beta})), depending on the width of
the film. Hitting the second interface there are again two options;
either the particle is refracted and leaves the system on the back
side of the film or it is reflected towards the first interface again,
but this time entering the other input port after acquiring another
phase shift on the way. Subsequently, a refraction leads to an exit on
the front side of the film and on reflection the particle is sent back
to the first input port of the second interface. This continues
until eventually, the particle leaves on the front or the back side of
the film.
The arrangement of the two DLMs is depicted in
Fig.~\ref{fig:film_dlm}. This setup corresponds to the behavior
described and shown in Fig.~\ref{fig:film}.

\subsection{Simulation results}
\begin{figure}
\begin{center}
\includegraphics[width=\textwidth]{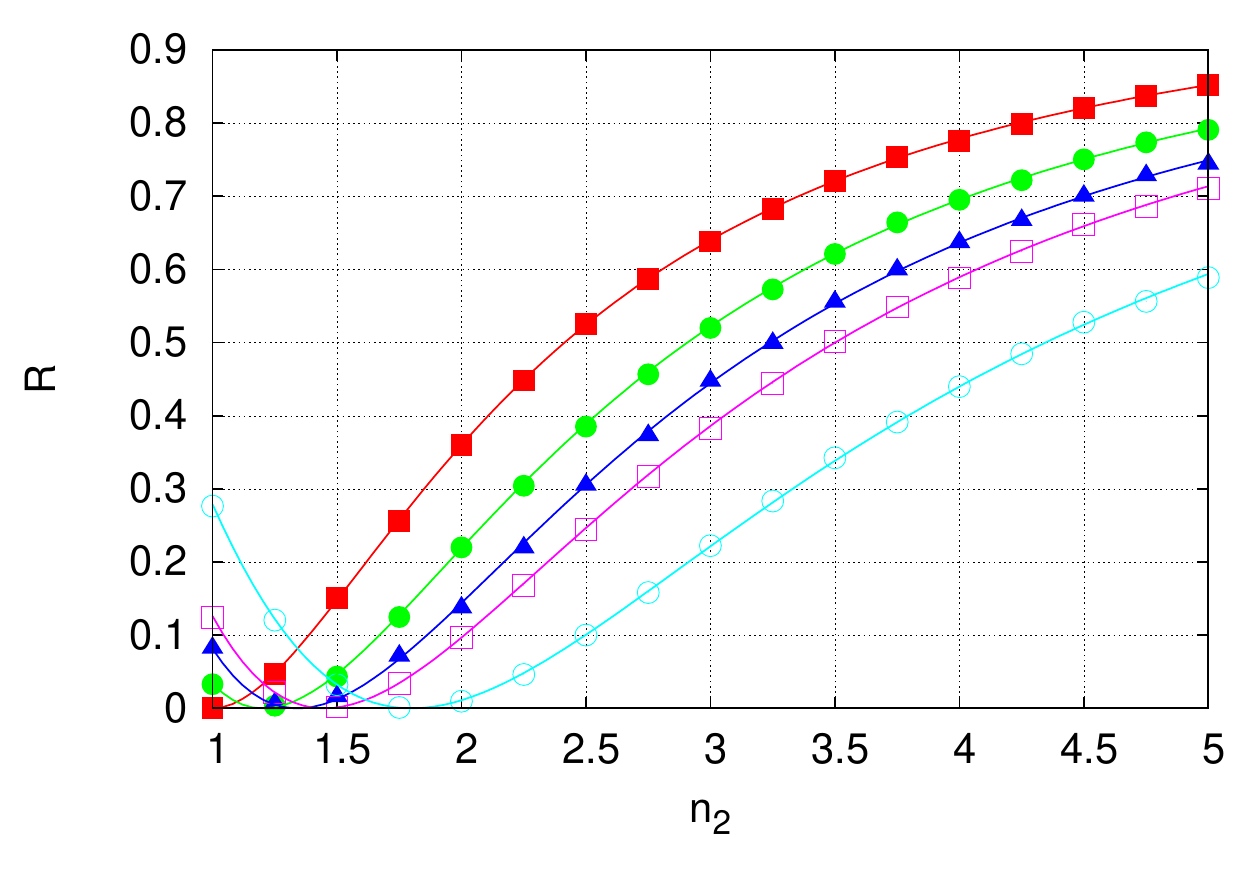}
\end{center}
\caption{Simulation results of the reflectivity $\mathcal{R}$
  of a quarter-wave plate, built from two plane parallel interfaces,
  for normal incidence and for various choices of refractive indices
  $n_1$, $n_2$ and $n_3$. Symbols represent simulation results with
  $\alpha=0.99$ and 100000 events per data point.
  Solid squares: $n_1=n_3=1$;
  Solid circles: $n_1=1$ and $n_3=1.45$;
  Triangles: $n_1=1$ and $n_3=1.8$;
  Open squares: $n_1=n_3=1.45$;
  Open circles: $n_1=n_3=1.8$;
  The solid lines are the wave theoretical predictions
  (Eq.~(\ref{eq:plate_reflectivity})).}
\label{fig:plate_lambda_fourth}
\end{figure}

\begin{figure}
\begin{center}
\includegraphics[width=\textwidth]{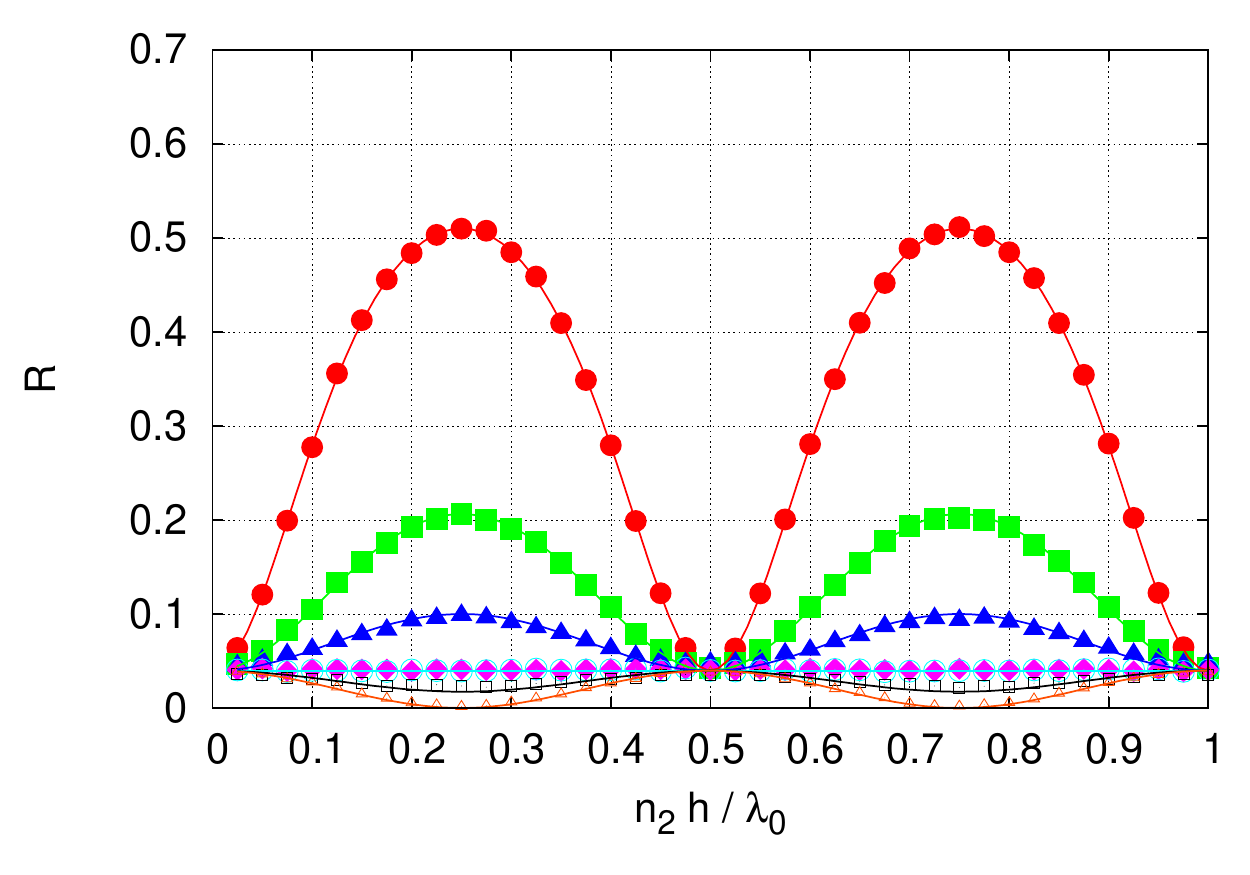}
\end{center}
\caption{Simulation results of the reflectivity $\mathcal{R}$ of a homogeneous dielectric film
  for normal incidence ($\theta_i=0$) in air ($n_1=n_3=1$) as a function of
  its thickness $h$ for various values of $n_2$.
  Solid circles: $n_2=3$;
  Solid squares: $n_2=2$;
  Triangles: $n_2=1.7$;
  Open circles: $n_2=1$;
  Diamonds: $n_2=1.5$;
  Open squares: $n_2=1.4$;
  Open triangles: $n_2=1.2$.
  The parameter $\alpha=0.99$ and 100000 events were processed for each set of parameters.
  The simulation results (markers) are in very good agreement with the wave
  theoretical expressions (solid lines, Eq.~(\ref{eq:plate_reflectivity})).}
\label{fig:plate_thickness}
\end{figure}

\begin{figure}
\begin{center}
\includegraphics[width=\textwidth]{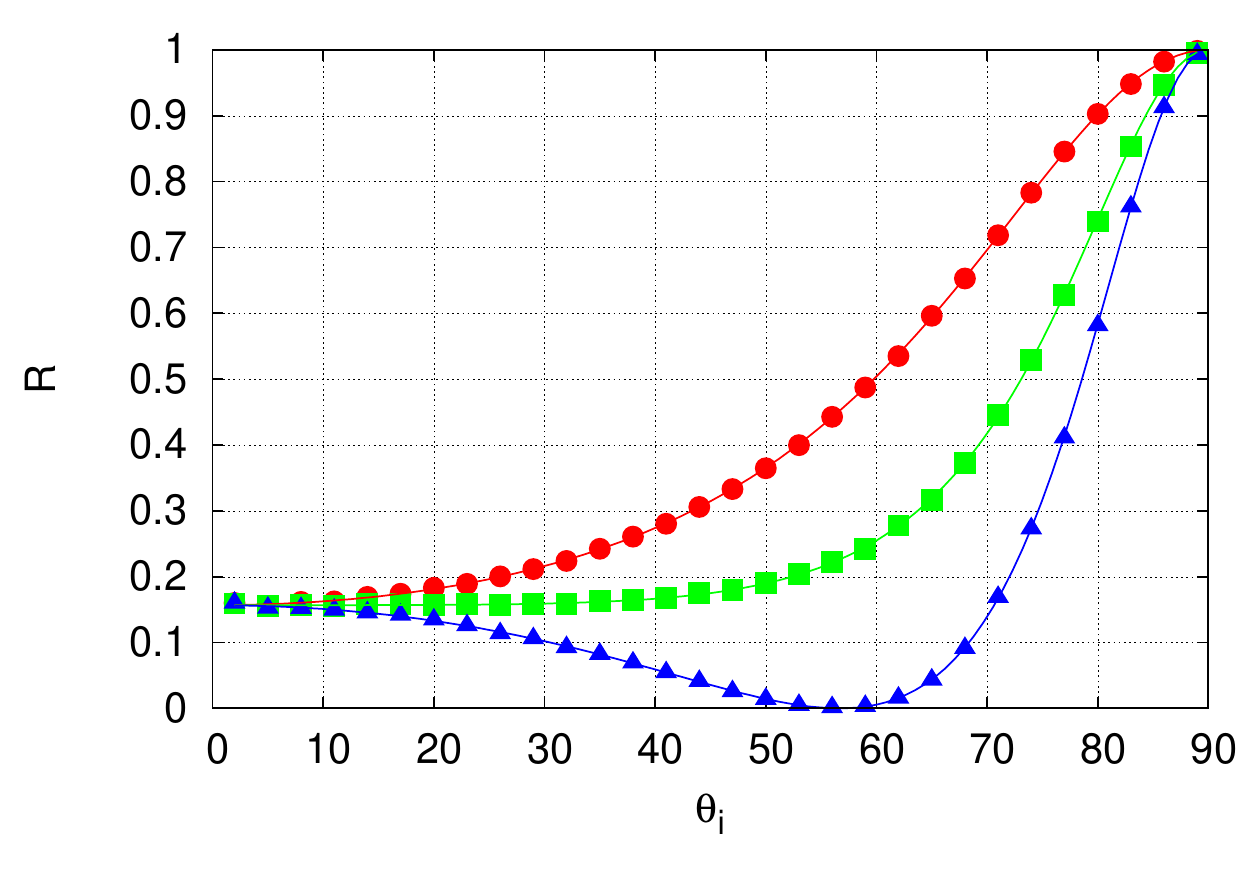}
\end{center}
\caption{Simulation data of the reflectivity $\mathcal{R}$ of a quarter wave plate as a function
  of the angle of incidence $\theta_i$ for different polarizations $\varpi$.
  Triangles: Polarization angle $\varpi=0$;
  Squares: $\varpi=\pi/4$;
  Circles: $\varpi=\pi/2$.
  The parameter $\alpha=0.99$ and 100000 events were processed for each data point.
  The simulation results (markers) are in very good agreement with the wave theoretical expressions
  (solid lines, Eq.~(\ref{eq:plate_reflectivity}))}
\label{fig:plate_angle}
\end{figure}

With the setup of Fig.~\ref{fig:film_dlm}, we simulate the behavior of homogeneous dielectric
films. The first analysis considers a plate of thickness $h=n_2\lambda_0/4$ (quarter-wave plate)
under normal incidence for various values of refractive indices. The results for
100000 events and $\alpha=0.99$, together with the results of Maxwell's theory, can be found in
Fig.~\ref{fig:plate_lambda_fourth}.
The values of the reflectivity predicted by  wave theory
(Eq.~\ref{eq:plate_reflectivity}) are reproduced with high precision
by the event-based simulation.
Figure~\ref{fig:plate_thickness} shows the reflectivity under normal
incidence of various plates surrounded by air ($n_1=n_3=1$) depending on the
thickness of the plate.
Again, the agreement with wave theory is very good. 
The variation with the incident angle for different polarizations is
shown in Fig.~\ref{fig:plate_angle}.
The wave theoretical predictions are all reproduced by the event-based
simulation, including features like zero reflectivity under the
Brewster angle for P-polarized ($\varpi=0$) light .

\subsection{Application: beam splitter}
\label{application_beamsplitter}
\begin{figure}
\begin{center}
\includegraphics[width=8cm]{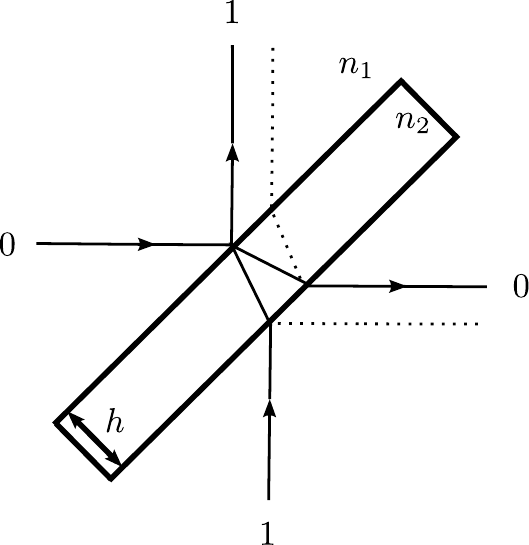}
\end{center}
\caption{Beam splitter built from two plane parallel interfaces. The
  incident angle is set to $45^\circ$, the width to
  $h=\lambda_0/4 n_2$, i.e.\ to an optical path
  length of a quarter wave and the refractive index $n_2$ is set to
  $n_2=1.86$, 
  so that we get a 50/50-beam splitter ($n_1=n_3=1$,
  $\varpi=\pi/2$). There are two input ports and two output
  ports. Due to the translational invariance of plane waves,
  parallel paths (indicated by dotted lines) can be matched. This is
  also true for paths emerging from multiple reflections within the
  plate (not shown here, compare Fig.~\ref{fig:film}).}
\label{fig:plate_beamsplitter}
\end{figure}

Having shown that the event-based model of the homogeneous dielectric film works as expected,
we now illustrate the modularity of the simulation approach  by building a 50/50-beam splitter.
The setup is shown in Fig.~\ref{fig:plate_beamsplitter}.
The incident angles are fixed to $45^\circ$, the width $h$ of the plate is
set to $h=\lambda_0/4 n_2$ and the refractive index
$n_2$ of the plate is set to $n_2=1.86$,
such that we get a 50/50-beam splitter for $\varpi=\pi/2$ (see
Eq.~(\ref{eq:plate_reflectivity})).

The incident particles on both input ports hit the beam splitter
such that the direction of the transmitted particle from one
port coincides with the direction of the reflected particle from
the other input port. Due to the translational invariance of plane
waves, parallel paths can be overlayed, as in the case of the plane parallel plate.

Next, we consider an experiment where we send, one-by-one,
particles carrying the phase information $\phi_0$ ($\phi_1$) to input port 0 (1).
The probability for a message to arrive on port 0 is $p_0$ and with probability of $1-p_0$ a message
arrives on port 1.
According to wave theory, the amplitudes $b_0$ and $b_1$ on the output ports 0 and
1 are given by~\cite{BORN64}
\begin{equation}
\left(\begin{array}{c}
b_0\\
b_1
\end{array}\right)
=\left(\begin{array}{cc}
t & r\\
r & t
\end{array}\right)
\left(\begin{array}{c}
\sqrt{p_0}e^{i\phi_0}\\
\sqrt{1-p_0}e^{i\phi_1}
\end{array}\right),
\end{equation}
with $r$ and $t$ given by
\begin{equation}
r=\frac{r_{12}+r_{23}e^{2i\beta}}{1+r_{12}r_{23}e^{2i\beta}},
\end{equation}
and
\begin{equation}
t=\frac{t_{12}t_{23}e^{i\beta}}{1+r_{12}r_{23}e^{2i\beta}}.
\end{equation}
The definitions of $r_{12}$ and $t_{12}$ are given in
Eqs.~(\ref{eq:plate_r}) and (\ref{eq:plate_t}), $r_{23}$ and $t_{23}$
are defined analogously.

Depending on the phase difference $\phi_0-\phi_1$ the
probability for a particle to exit on the output port 0 is given by
\begin{equation}
|b_0|^2=\frac{1}{2}+\sqrt{p_0(1-p_0)}\sin(\phi_0-\phi_1).
\end{equation}

\begin{figure}
\begin{center}
\includegraphics[width=\textwidth]{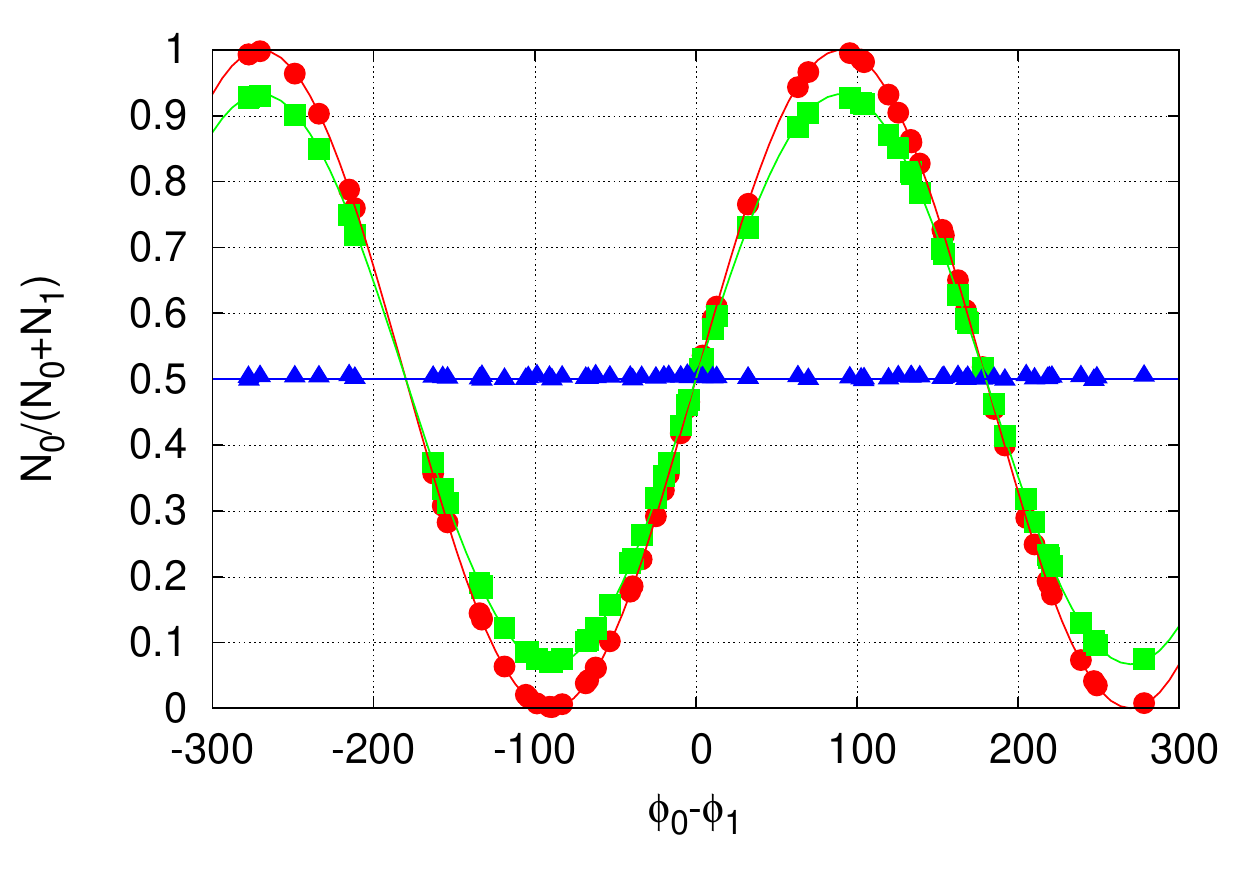}
\end{center}
\caption{Simulation results for a 50/50-beam splitter built from two plane
  interfaces. The parameter $\alpha=0.98$ and
  100000 events have been processed for each data point. The phases
  $\phi_0$ and $\phi_1$ were drawn from uniform random
  distributions. We measured the normalized intensity $N_0/(N_0+N_1)$,
  with $N_0$ and $N_1$ being the number of events on the corresponding
  output port. Depending on the probability $p_0$ of an event
  entering on input port 0, we get different results which agree
  very well with the wave theoretical predictions (solid lines).
  Triangles: $p_0=1$;
  Circles: $p_0=0.5$;
  Squares: $p_0=0.25$.
  }
\label{fig:beamsplitter_0.98}
\end{figure}

We have run the single-event simulation with $\alpha=0.98$ and
$N_0+N_1=100000$ events for each value of the phase difference, where
$N_0$ and $N_1$ denote the number of events on the output port 0
and 1, respectively. We determined the normalized intensity
$N_0/(N_0+N_1)$ detected on output port 0. The results for various
values of $p_0$ are shown in Fig.~\ref{fig:beamsplitter_0.98}.
The event-based simulation is in very good agreement with the
behavior predicted by wave theory.

\subsection{Discussion}
We have shown that we can use an event-based simulation approach to
describe the behavior of a homogeneous dielectric film. With the
proper choice of parameters we can use this as a beam splitter. This
beam splitter can be used as building block for optics experiments like
the Mach-Zehnder interferometer (see section~\ref{mach-zehnder}).

\section{Light propagation through a periodic multilayer}
\label{multilayer}
\subsection{Wave theory}
\begin{figure}
\begin{center}
\includegraphics[width=12cm]{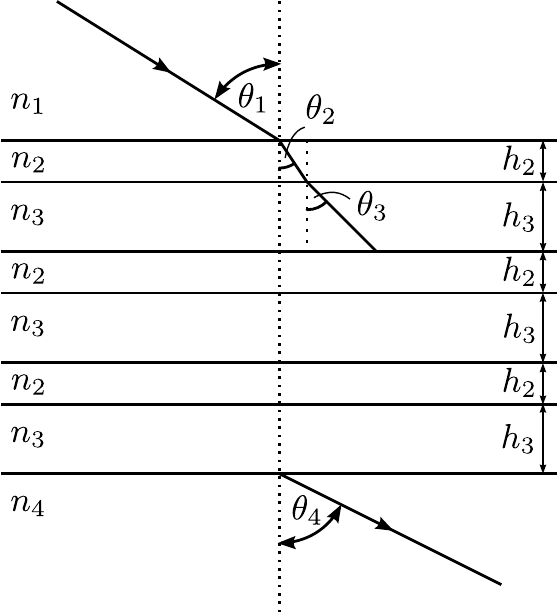}
\end{center}
\caption{Scheme of a periodic multilayer. It consists of a stack of
  layers with alternating refractive indices, $n_2$ and $n_3$, and
  thicknesses, $h_2$ and $h_3$ between two media with refractive
  indices $n_1$ and $n_4$. Drawn here are only three periods, but the
  number of periods can be chosen arbitrarily. It is also possible to
  end with the transition from $n_2$ to $n_4$ skipping the last layer
  with refractive index $n_3$.}
\label{fig:periodic_multilayer_scheme}
\end{figure}

A periodic multilayer consists of a succession of homogeneous layers
of alternating refractive indices, $n_2$ and $n_3$, and thicknesses,
$h_2$ and $h_3$, between two homogeneous media with refractive indices
$n_1$ and $n_4$ (see Fig.~\ref{fig:periodic_multilayer_scheme}).
For quarter wave layers ($n_2h_2=n_3h_3=\lambda_0/4$) at normal
incidence the reflectivity for a total of $N$ interfaces is given by~\cite{BORN64} 
\begin{equation}
\mathcal{R}_N=\left(\frac{1-\left(\frac{n_4}{n_1}\right)\left(\frac{n_2}{n_3}\right)^{N-1}}{1+\left(\frac{n_4}{n_1}\right)\left(\frac{n_2}{n_3}\right)^{N-1}}\right)^2,\quad\text{if
  $N$ is odd},
\label{eq:multilayer_R_odd}
\end{equation}
i.e., if the stack ends on $n_3\rightarrow n_4$, and
\begin{equation}
\mathcal{R}_N=\left(\frac{1-\left(\frac{n_2}{n_1}\right)\left(\frac{n_2}{n_4}\right)\left(\frac{n_2}{n_3}\right)^{N-2}}{1+\left(\frac{n_2}{n_1}\right)\left(\frac{n_2}{n_4}\right)\left(\frac{n_2}{n_3}\right)^{N-2}}\right)^2,\quad\text{if
  $N$ is even},
\label{eq:multilayer_R_even}
\end{equation}
i.e., if the stack ends on $n_2\rightarrow n_4$.

\subsection{Event-based simulation}
\begin{figure}
\begin{center}
\includegraphics{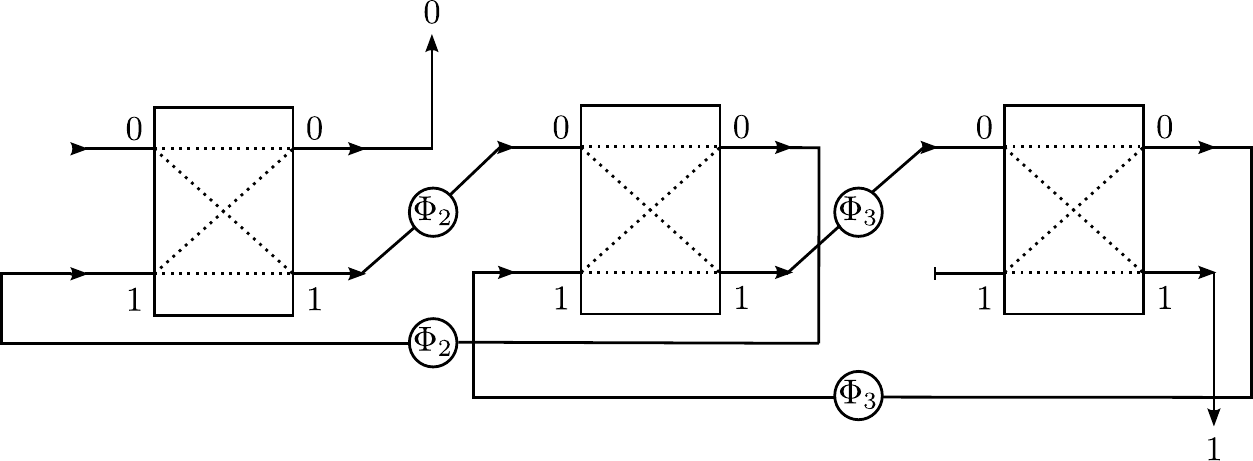}
\end{center}
\caption{Network of DLMs for the simulation of a periodic
  multilayer. Here, only a single period ($n_1\rightarrow
  n_2\rightarrow n_3\rightarrow n_4$) with 3 interfaces is shown, but
  in principle it can be generalized to arbitrarily many periods. The
  first DLM on the left simulates a plane interface between media with
  refractive indices $n_1$ and $n_2$. The second DLM describes the
  transition from $n_2$ to $n_3$ and the third DLM simulates the
  interface between media with refractive indices $n_3$ and $n_4$. By
  inserting more DLMs before the rightmost one, it is possible to
  build any multilayer. For our simulation, we chose a periodic setup
  with alternating layers with reflective indices $n_2$ and $n_3$.}
\label{fig:periodic_multilayer_dlms}
\end{figure}

The method for simulating a plate consisting of two interfaces
(section~\ref{plate}) can be generalized to the case of a
multilayer. In this case we concatenate $N$ DLMs, each one simulating
the behavior of a single interface. The schematic diagram of the DLM network is depicted
in Fig.~\ref{fig:periodic_multilayer_dlms}.
A message, that is sent into the network can propagate back and forth
through any DLM, until eventually it leaves the network through
port 0 of the first DLM or port 1 of the last DLM. This
corresponds to multiple transmissions and reflections of a photon
within the multilayer, until eventually it exits on the front or
the back side.

\subsection{Simulation results}
\begin{figure}
\begin{center}
\includegraphics[width=\textwidth]{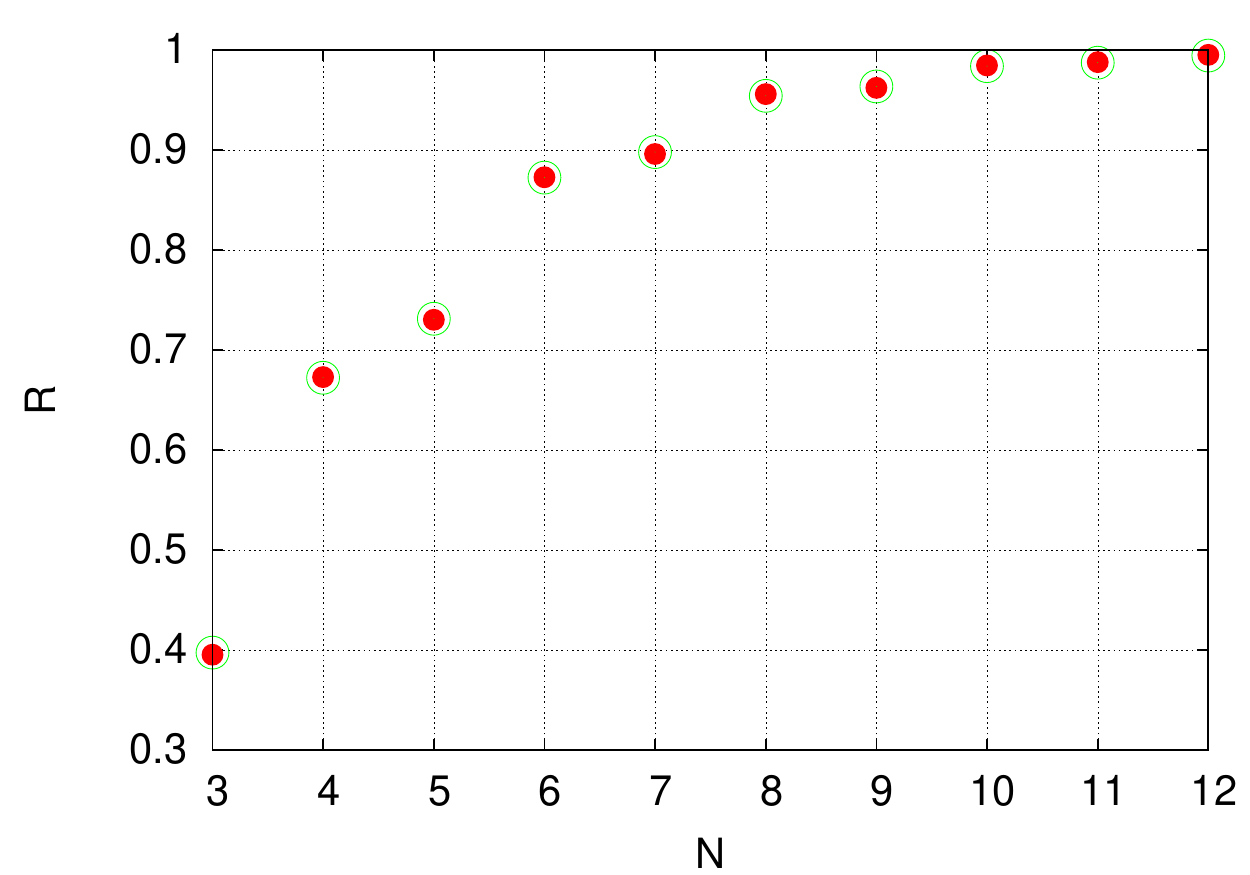}
\end{center}
\caption{Reflectivity $\mathcal{R}$ of a periodic multilayer as a function
  of the number of interfaces $N$.
  Closed circles: Event-based simulation data.
  Open circles: Wave theory (Eqs.~(\ref{eq:multilayer_R_odd}) and
  (\ref{eq:multilayer_R_even})).
  The parameters were chosen to resemble
  an experiment with quarter wave films of zinc sulphide and cryolite
  at normal incidence~\cite{BORN64}: $n_1=1$, $n_2=2.3$, $n_3=1.35$,
  $n_4=1.52$, $n_2h_2=n_3h_3=\lambda_0/4$, $\theta_i=0$. The
  simulation was carried out with $\alpha=0.998$ and 1000000 events
  per data point. The results of our event-based approach agree
  very well with the predictions of wave theory.}
\label{fig:multilayer_N}
\end{figure}

\begin{figure}
\begin{center}
\includegraphics[width=\textwidth]{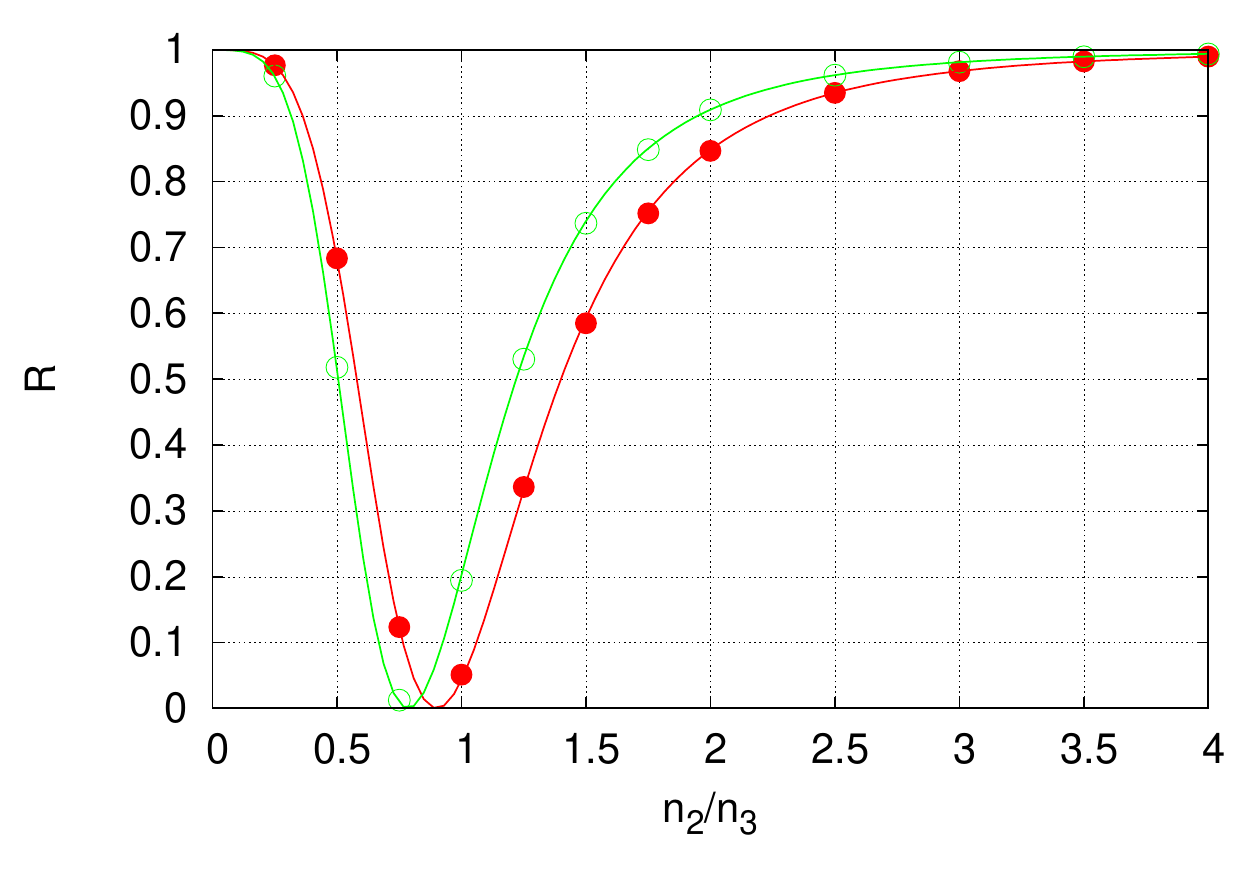}
\end{center}
\caption{Reflectivity $\mathcal{R}$ of a periodic multilayer depending
on the ratio $n_2/n_3$ of the refractive indices for $N=5$ (solid circles) and $N=6$ (open circles)
interfaces. Data points are simulation results as obtained  with
$\alpha=0.99$ and 100000 events for each point. Solid lines are
results derived from wave theory (Eqs.~(\ref{eq:multilayer_R_odd})
and (\ref{eq:multilayer_R_even})).
Model parameters are: $n_1=1$, $n_2=2$,
$n_4=1.52$, $n_2h_2=n_3h_3=\lambda_0/4$, $\theta_i=0$, with $n_3$ varying.
The event-by-event simulation gives the same results as the
wave theoretical description (solid lines).}
\label{fig:multilayer_n}
\end{figure}

We study the case of a periodic multilayer. The sequence of refractive
indices along the stack is $n_1$, $n_2$, $n_3$, $n_2$, $n_3$,
$\ldots$, $n_2$, ($n_3$), $n_4$
(figure~\ref{fig:periodic_multilayer_scheme}).
Figure~\ref{fig:multilayer_N} shows the reflectivity (normalized
intensity on output port 0) depending on the number of interfaces.
Another analysis shows the reflectivity depending on the ratio
$n_2/n_3$ (Fig.~\ref{fig:multilayer_n}).

\subsection{Discussion}
We have shown that the approach introduced in section~\ref{interface}
can be generalized to multilayers by concatenating multiple DLMs. The
basic building block is a DLM simulating a plane interface between two
homogeneous dielectric media. A network of multiple DLMs can be used
to form more complex structures. We studied the case of a periodic
multilayer and found excellent agreement of the event-based simulation
results with the wave theoretical description.

\section{Event-based simulation of a Mach-Zehnder interferometer}
\label{mach-zehnder}
\subsection{Wave theory}
\begin{figure}
\begin{center}
\includegraphics[width=\textwidth]{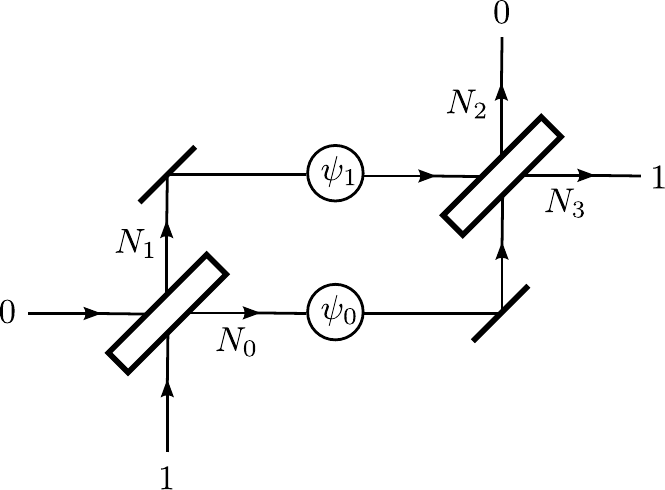}
\end{center}
\caption{Schematic setup of a Mach-Zehnder interferometer, consisting
of two beam splitters and two devices that perform a specific phase
shift. A photon is sent on one of the input ports and leaves the
beam splitter on one of its output ports. Depending on which path
of the interferometer a photon takes, its phase is changed by
either $\psi_0$ or $\psi_1$. It then interacts with the second beam
splitter and eventually, a photon is detected on one of the output
ports of the interferometer. In the simulation we denote the
photon numbers on each path with $N_0$, $N_1$, $N_2$, and $N_3$.
The counts $N_0$ and $N_1$ give the number of particles
that travel from the first to the second beamsplitter via
one of the two pathways (but never via the two pathways simultaneously)}.

\label{fig:mach-zehnder_setup}
\end{figure}

The Mach-Zehnder interferometer is a device that is sensitive to
relative phase shifts~\cite{BORN64}. The schematic setup is depicted in
Fig.~\ref{fig:mach-zehnder_setup}.
At any time, there is at most a single photon traveling through the
system. After passing the first beam splitter the photon gets an
additional phase shift depending on the path that it follows. It
passes the second beam splitter and is detected on one of the output
ports.
According to wave theory the amplitudes of the photons $(b_0, b_1)$ in
the output ports 0 ($N_2$) and 1 ($N_3$) are given by
\begin{equation}
\left(\begin{array}{c}
b_0\\
b_1
\end{array}\right)
=\frac{1}{2}
\left(\begin{array}{cc}
1 & i\\
i & 1
\end{array}\right)
\left(\begin{array}{cc}
e^{i\psi_0} & 0\\
0 & e^{i\psi_1}
\end{array}\right)
\left(\begin{array}{cc}
1 & i\\
i & 1
\end{array}\right)
\left(\begin{array}{c}
a_0\\
a_1
\end{array}\right),
\end{equation}
with $(a_0, a_1)$ being the amplitudes in the input ports and
$\psi_0$, $\psi_1$ describing the additional phase rotations in the
specific arm of the interferometer.

For input on port 0 only, i.e.\ $(a_0,
a_1)=(\cos\phi_0+i\sin\phi_0, 0)$, we get the probability distribution
\begin{equation}
|b_0|^2=\sin^2\left(\frac{\psi_0-\psi_1}{2}\right),\quad
|b_1|^2=\cos^2\left(\frac{\psi_0-\psi_1}{2}\right).
\end{equation}

\subsection{Event-based simulation}
We simulate a Mach-Zehnder interferometer with our event-based
approach by using the same building blocks as shown in
Fig.~\ref{fig:mach-zehnder_setup}; we use two beam splitters, two
phase shifters, one single-photon source (not shown) and two detectors (not shown).
The beam splitters are built from more fundamental single interfaces as described in
section~\ref{application_beamsplitter}. The communication between the
components of the interferometer is mediated by single photons
only. Each photon carries a message that is just its phase information.

\begin{figure}
\begin{center}
\includegraphics[width=\textwidth]{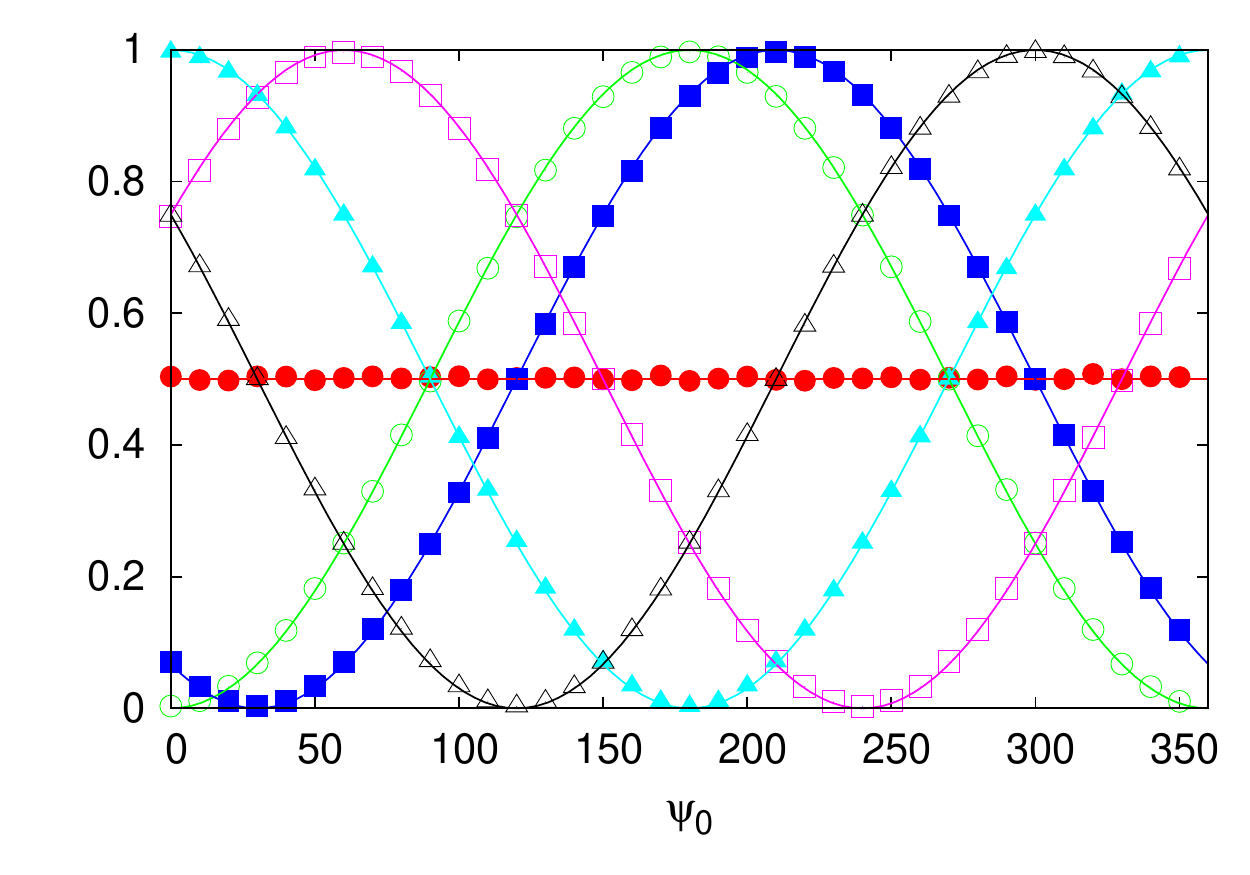}
\end{center}
\caption{Normalized intensities in Mach-Zehnder interferometer (see
  Fig.~\ref{fig:mach-zehnder_setup}) depending on the phase shift
  $\psi_0$ for various values of the phase shift $\psi_1$. The
  data points are simulation results using 100000 events and
  $\alpha=0.98$.
  Solid circles: $N_0/(N_0+N_1)$ for $\psi_1=0$;
  Open circles: $N_2/(N_2+N_3)$ for $\psi_1=0$;
  Solid squares: $N_2/(N_2+N_3)$ for $\psi_1=30^\circ$;
  Open squares: $N_2/(N_2+N_3)$ for $\psi_1=240^\circ$;
  Solid triangles: $N_3/(N_2+N_3)$ for $\psi_1=0$;
  Open triangles: $N_3/(N_2+N_3)$ for $\psi_1=300^\circ$;
  The solid lines are the predictions of wave theory.}
\label{fig:mach-zehnder_results}
\end{figure}

The simulation results for input on a single port only, i.e.\
$(\cos\phi, \sin\phi)$ on port 0 and no input on port 1, are
shown in Fig.~\ref{fig:mach-zehnder_results}. The phase $\phi$
is chosen randomly but fixed from an equal distribution
in the range $[0, 360^\circ]$.
Each data point represents 100000 events $(N_0+N_1=N_2+N_3=100000)$
and $\alpha$ is set to $\alpha=0.98$. The angle of rotation $\psi_0$
is varying in steps of $10^\circ$ and the normalized intensities for
the various ports are determined from the event-based simulation.

The agreement between the event-based simulation and the wave
theoretical description is very good if $\alpha$ is close to 1. This has
been shown for event-based simulations with a DLM model that simulates
a beam splitter as a whole~\cite{RAED05b,RAED05c,RAED05d}, not as
a collection of interfaces, as is done in the present work.
Thus, we have shown that DLMs describing beam splitters can be built up from
more fundamental building blocks, namely single interfaces, illustrating
the modularity of our simulation method.

\subsection{Discussion}

Our event-based simulation of a MZI shows that this optics experiment
can be simulated by building a network of a
basic building block, the DLM-based machine that simulates an interfaces between two
homogeneous media.
The simulation results for the basic building block and the more complicated
networks such as a plane parallel plate, a beam splitter, a periodic stratified medium,
and a MZI are all in excellent agreement with the corresponding results of Maxwell's theory,
demonstrating that our event-based approach is modular.
Crucial for the DLM-based simulation to yield the results of Maxwell's theory
is that the parameter $\alpha$ which controls the dynamics of the DLM is close to one~\cite{RAED05b,RAED05c,RAED05d}.

\section{Conclusion}
\label{conclusion}
We have shown that basic optical phenomena such as reflection, refraction and interference
can be simulated by an event-based, particle-like approach.
Our computational approach has the following features:

\begin{itemize}
\item{It yields the stationary solution of the Maxwell equations
by simulating particle trajectories only.}
\item{Material objects are represented by DLM-based units placed on a boundary
of these objects, which in practice involves some form of discretization. Apart from this discretization,
all calculations are performed using Euclidean geometry.
}
\item{Unlike wave equation solvers, it does not suffer
from artifacts due to the unavoidable termination of the simulation volume~\cite{TAFL05}: Particles that
leave this volume can simply be removed from the simulation.}
\item{Unlike wave equation solvers which may consume substantial computational resources (i.e. memory and CPU time)
to simulate the propagation of waves in free space, it calculates the motion of the corresponding particles in free space
at almost no computational cost.}
\item{Modularity: Starting from the unit that simulates the behavior of a
plane interface between two homogeneous media other optical components
can be constructed by repeated use of the same unit.}
\end{itemize}

We believe that the work presented in this paper may open a route to
rigorously include the effects of interference in ray-tracing software.
For this purpose, it is necessary to extend the DLM-based model
for lossless dielectric materials to, say a Lorentz model for the response of material
to the electromagnetic field~\cite{TAFL05}.
This may be done by simple modifications of the DLM update rule,
an extension that we leave for future research.


\section*{References}



\bibliographystyle{elsarticle-num}
\bibliography{../../epr}







\end{document}